\documentclass[prb,preprint]{revtex4-1} 

\usepackage[colorlinks,citecolor=blue,urlcolor=blue,linkcolor=blue]{hyperref}
\usepackage{amsmath}  
\usepackage{amsfonts} 
\usepackage{graphicx} 

\begin{document}


\title{Tesla's fluidic diode and the electronic-hydraulic analogy}

\author{Quynh M. Nguyen}
\altaffiliation[]{Department of Physics, New York University, New York, NY 10003} 

\author{Dean Huang}

\author{Evan Zauderer}

\author{Genevieve Romanelli}
\author{Charlotte L. Meyer}
\author{Leif Ristroph}
\email{Email: ristroph@cims.nyu.edu}

\affiliation{Applied Math Lab, Courant Institute of Mathematical Sciences, New York University, New York, NY 10012}


\date[ ]{\today}

\begin{abstract}
Reasoning by analogy is powerful in physics for students and researchers alike, a case in point being electronics and hydraulics as analogous studies of electric currents and fluid flows. Around 100 years ago, Nikola Tesla proposed a flow control device intended to operate similarly to an electronic diode, allowing fluid to pass easily in one direction but providing high resistance in reverse. Here we use experimental tests of Tesla’s diode to illustrate principles of the electronic-hydraulic analogy. We design and construct a differential pressure chamber (akin to a battery) that is used to measure flow rate (current) and thus resistance of a given pipe or channel (circuit element). Our results prove the validity of Tesla’s device, whose anisotropic resistance derives from its asymmetric internal geometry interacting with high-inertia flows, as quantified by the Reynolds number (here, $\textrm{Re} \sim 10^3$). Through the design and testing of new fluidic diodes, we explore the limitations of the analogy and the challenges of shape optimization in fluid mechanics. We also provide materials  that may  be  incorporated  into  lesson plans  for  fluid  dynamics  courses, laboratory modules and further research projects.
\end{abstract}

\maketitle 

\section{Introduction}

Nikola Tesla is celebrated for his creativity and ingenuity in electricity and magnetism. Perhaps part of his genius lies in connecting ideas and concepts that do not at first glance appear related, and Tesla's writings and record of inventions suggest he reasoned by analogy quite fluidly. While he is best known for inventing the AC motor -- which transforms oscillating electric current into one-way mechanical motion -- Tesla also invented a lesser known device intended to convert oscillating fluid flows into one-way flows. Just around 100 years ago, and while living in New York City not far from our Applied Math Lab, Tesla patented what he termed a \textit{valvular conduit}, \cite{patent} as shown in Fig. \ref{fig:patent}. The heart of the device is a channel through which a fluid such as water or air can pass, and whose intricate and asymmetric internal geometry is intended to present strongly different resistances to flow in one direction versus the opposite direction. From his writing in the patent, Tesla has clear purposes for the device: To transform oscillations or pulsations, driven perhaps by a vibrating piston, to one-way motion either of the fluid itself (the whole system thus acting as a pump or compressor) or of a rotating mechanical component (\textit{i.e.} a rotary motor).

Whether called pumping, valving, rectification or AC-to-DC conversion, this operation is one example of the analogy between electrodynamics and hydrodynamics. And the electronic-hydraulic analogy is but one of many such parallels that show up in physics and across all fields of science and engineering. As emphasized by eminent physicists such as Maxwell and Feynman, reasoning by analogy is one of the powerful tools that allow scientists, having understood one system, to quickly make progress in understanding others. \cite{feynmanvol2,maxwell,pask2003mathematics} It is also a valuable tool in teaching challenging scientific concepts.\cite{duit1991role} Here, we explore this style of reasoning in the context of Tesla's invention, whose operation is analogous to what we now call an electronic \textit{diode}. In hydraulic terms, this device plays the role of a \textit{check valve}, which typically involves an internal moving element such as a ball to block the conduit against flow in the reverse direction. The practical appeal of Tesla's diode, in addition to its pedagogical value,\cite{stith2019tesla} is that it involves no moving parts and thus no components that wear, fail or need replacement.\cite{patent} 

\begin{figure}
\centering
\includegraphics[width=16cm]{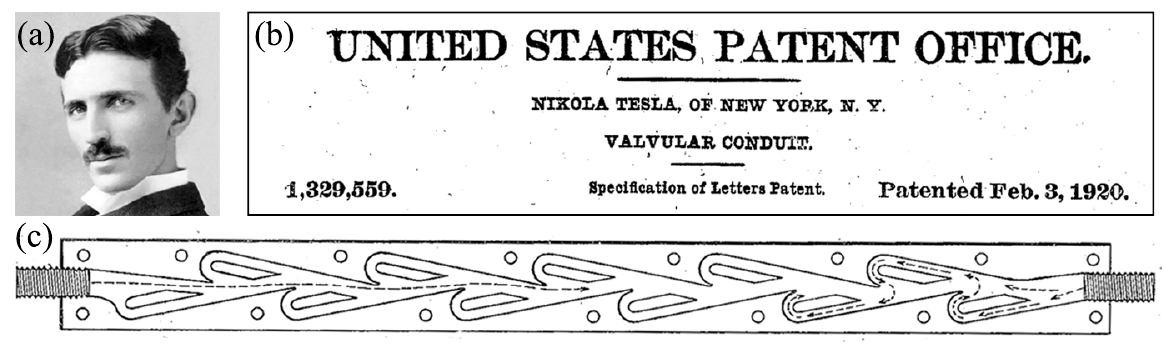}\vspace{-0.6cm}
\caption{Nikola Tesla's valvular conduit. (a) The inventor Nikola Tesla (1856-1943). (b) Title to Tesla's 1920 patent for the valvular conduit \cite{patent}. (c) Schematic showing the conduit's internal geometry. The channel is intended to allow fluid to flow from left to right (forward direction) with minimal resistance while providing large resistance to reverse flow. }
\label{fig:patent}
\end{figure}

In exploring the electronic-hydraulic analogy through Tesla's diode, we also provide pedagogical materials that may be incorporated into lesson plans for fluid dynamics courses and especially laboratory courses. For students of physics, electricity and electronics are often motivated by hydraulics,\cite{smith1974,bauman1980,greenslade2003,pfister2004glitter,pfister2014sponge} in which voltage is akin to pressure, current to flow rate, electrical resistance in a wire to fluidic resistance of a pipe, etc. However, in the modern physics curriculum, electrodynamics quickly outpaces hydrodynamics, and most students have better intuition for voltage than pressure! Lesson plans and laboratory modules based on this work would have the analogy work in reverse, \textit{i.e.} employ ideas from electronics to guide reasoning about Tesla's device, and thereby learn the basics of \textit{fluidics} or the control of fluid flows. As such, we present an experimental protocol for testing Tesla's diode practically, efficiently and inexpensively while also emphasizing accuracy of measurements and reproducibility of results. The feasibility of our protocol is vetted through the direct participation of undergraduate (DH and EZ) and high school (GR and CM) students in this research. We also suggest and explore avenues for further research, such as designing and testing new types of fluidic diodes.

\section{Tesla's device, proposed mechanism, efficacy and utility}

Tesla's patent is an engaging account of the motivations behind the device, its design, proposed mechanism and potential uses. In this section, we briefly summarize the patent and highlight some key points, quoting Tesla's words wherever possible. \cite{patent} The general application is towards a broad class of machinery in which ``fluid impulses are made to pass, more or less freely, through suitable channels or conduits in one direction while their return is effectively checked or entirely prevented.'' Conventional forms of such valves rely on ``carefully fitted members the precise relative movements of which are essential'' and any mechanical wear undermines their effectiveness. They also fail ``when the impulses are extremely sudden or rapid in succession and the fluid is highly heated or corrosive.'' Tesla aims to overcome these shortcomings through a device that carries out ``valvular action... without the use of moving parts.'' The key is an intricate but static internal geometry consisting of ``enlargements, recesses, projections, baffles or buckets which, while offering virtually no resistance to the passage of the fluid in one direction, other than surface friction, constitute an almost impassable barrier to its flow in the opposite sense." Figure \ref{fig:patent}(c) is a view of the channel internal geometry, where the fluid occupies the central corridor and the eleven ``buckets'' around the staggered array of solid partitions.

Without making any concrete claims to having investigated its mechanism, Tesla relates the function of the device to the character of the flows generated within the channel, as indicated by dashed arrows in Fig. \ref{fig:patent}(c). In the left-to-right or forward direction, the flow path is ``nearly straight''. However, if driven in reverse, the flow will ``not be smooth and continuous, but intermittent, the fluid being quickly deflected and reversed in direction, set in whirling motion, brought to rest and again accelerated.'' The high resistance is ascribed to these ``violent surges and eddies'' and especially the ``deviation through an angle of $180^\circ$'' of the flow around each ``bucket'' and ``another change of $180^\circ$... in each of the spaces between two adjacent buckets,'' as indicated by the arrows on the right of Fig. \ref{fig:patent}(c).

The effectiveness of the device can be quantified as ``the ratio of the two resistances offered to disturbed and undisturbed flow.'' Without directly stating that he constructed and tested the device, Tesla repeats a claim about its efficacy: ``The theoretical value of this ratio may be 200 or more;'' ``a coefficient approximating 200 can be obtained;'' and ``the resistance in reverse may be 200 times that in the normal direction.'' The experiments described below will directly assess this effectiveness or \textit{diodicity}.\cite{forster1995}

Much of the remaining portions of Tesla's patent are devoted to example uses. The first is towards the ``application of the device to a fluid propelling machine, such as, a reciprocating pump or compressor.'' The second is intended to drive ``a fluid propelled rotary engine or turbine.'' The explanations and accompanying diagrams are quite involved, and at the end of this paper we offer our own applications that we think capture Tesla's intent in simpler contexts.

\section{Experimental method to test Tesla's diode}

\begin{figure}
\centering
\includegraphics[width=15cm]{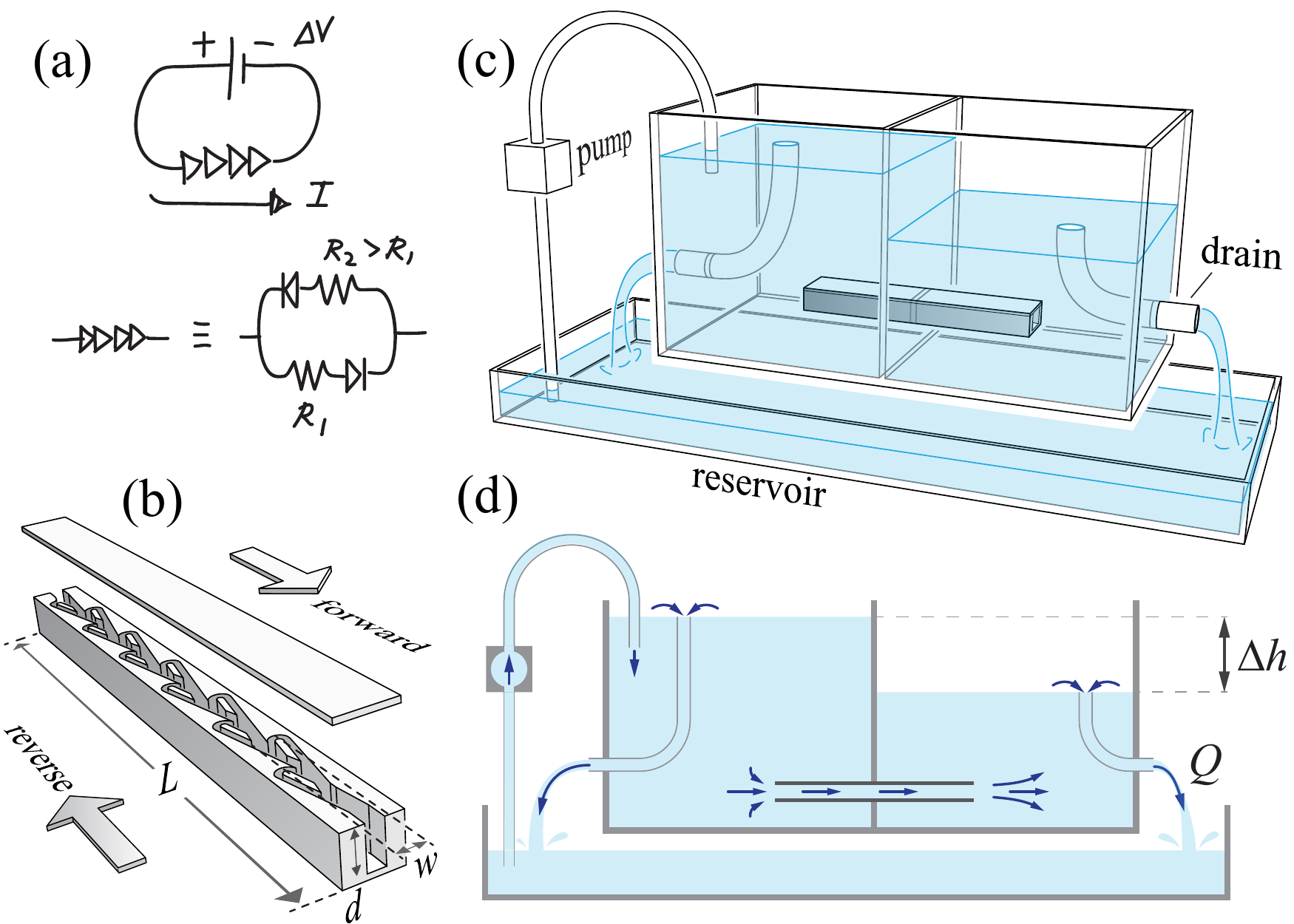}\vspace{-0.6cm}
\caption{Experimental apparatus for testing Tesla's diode. (a) Circuit analogs. The resistance of an unknown component is characterized by imposing a voltage difference and measuring current. A `leaky' diode may be represented in terms of resistors and ideal diodes. (b) Rendering of Tesla's conduit that can be 3D printed or laser cut. Relevant dimensions include total length $L$, depth or height $d$, and average wetted width $w$ (c) Perspective view of experimental apparatus. Two upper chambers are connected only via the conduit to be tested, and both overflow through internal drains to a lower reservoir. The water level of the higher is maintained by an overspill mechanism with a pump. Flow through the channel into the lower chamber induces overflow into the reservoir. (d) Schematic of pressure source chamber. The height difference $\Delta h$ is set by variable-height drains, and the resulting volume flow rate $Q$ is measured. }
\label{fig:setup}
\end{figure}

It is unclear whether Tesla ever constructed and tested a prototype, and this fact is obscured by the vague language used in the patent.\cite{patent} In any case, no data are provided. In the 100 years since, there has been much research into modified versions of Tesla's conduit \cite{paul,bardell2000,nostaggertesla1, nostaggertesla2,ziczactesla,3dtesla,thompson,truong,zhang,mohammad,nobakht,gamboa,morris,forster2002,forster2007,anag,ansari,vries,reconstruct,lin2015,deng,thompson2013transitional} and asymmetric channels generally for micro- and macro-fluidic applications.\cite{forster1995,stemme,microbucket,micro-sawtooth,ratchet,topodesign,nguyen,sousa,fadl,chen} However, to our knowledge, there are no studies on conduits faithful to Tesla's original geometry. Here we use modern rapid prototyping techniques to manufacture such a channel, and we outline an experimental characterization of its hydraulic resistance that uses everyday instruments like rulers, beakers and stopwatches to yield high-precision measurements.

We start with motivation from the electronic-hydraulic analogy. Suppose we have a circuit element of unknown and possibly anisotropic resistance, which we give the symbol of four arrowheads in Fig. \ref{fig:setup}(a). To characterize the element, we wish to impose a voltage difference $\Delta V$ using a battery or voltage source and measure the resulting current $I$, perhaps using an ammeter (not shown). The resistance is then given by Ohm's law $R=\Delta V / I$. Following the usual analogy, we wish to impose a pressure difference $\Delta p$ across the conduit, measure the resulting volume flow rate $Q$ and thus infer the resistance $R=\Delta p / Q$.

The conduit plays the role of the unknown element and is readily manufactured thanks to the modern convenience of rapid prototyping. We first digitize the channel geometry directly from the patent to arrive at a vector graphics file, and we have tested both 3D-printed and laser-cut realizations. Having achieved highly reproducible results on the latter, here we report on a design cut from clear acrylic sheet, a rendering of which is shown in Fig. \ref{fig:setup}(b). The channel tested has height or depth $d = 1.9$ cm, length $L = 30$ cm and average wetted width $w=0.9$ cm, and its planform geometry faithfully reflects Tesla's design. Gluing a top using acrylic solvent ensures a waterproof seal. If the desired channel is deeper than the maximum thickness permitted for a given laser-cutter, several copies may be cut and bonded together in a stack. Channels may also be 3D-printed, in which case waterproofing can be achieved by painting the interior with acrylic solvent to seal gaps between printed layers.

What serves the function of a hydraulic battery? We desire a means for producing a pressure difference across the channel, thereby driving a flow, and it is natural to employ columns of water as sources of hydrostatic pressure. If the channel bridges two columns of different heights, water flows through the channel from the higher to the lower. The challenge is to achieve an \textit{ideal pressure source} (akin to an ideal voltage source) that maintains the heights and thus pressures even as flows drain from one and into the other. This is accomplished using overflow mechanisms, as detailed in the experimental apparatus of Fig. \ref{fig:setup}(c). Two water-filled chambers are connected only by the channel being tested. Each chamber has an internal drain that can be precisely positioned vertically via a translation stage. The heights of the drains set the water levels and thus the flow direction, which can be reversed by switching which drain is higher. The draining of the high side through the channel is compensated by a pump that draws from a reservoir; the pump is always run sufficiently fast so as to just overflow the chamber and thus maintain its level. The lower side is fed only by the flow from the channel, and hence the flow through its drain and out the side to the reservoir represents the flow through the channel itself. The whole system is closed and can be run indefinitely.

A sectional view of the apparatus is shown in Fig. \ref{fig:setup}(d). A desired difference in water heights $\Delta h$ can be obtained by adjusting the drain heights. Considering hydrostatic pressure, we argue that the pressure difference across the channel is $\Delta p = \rho g \Delta h$, where $\rho = 1.0 ~\textrm{g}/\textrm{cm}^3$ is the density of water and $g = 980 ~\textrm{cm}/\textrm{s}^2$ is gravitational acceleration. (We use the centimeter-gram-second or CGS system of units throughout this work, as it proves convenient given the experimental scales.) A more thorough analysis of the pressures is detailed in Section VI. The resulting volume rate $Q$ of water flowing from high to low through the channel can then be measured by timing with a stopwatch the filling of a beaker of known volume that intercepts the flow exiting from the lower chamber. Large vessels and consequently long measurement times ensure highly accurate results, and measurements may be repeated to ensure reproducibility. Resistance can then be calculated as $R=\Delta p / Q$. Importantly, the experimental scales and working fluid are chosen to achieve strongly inertial flows, as quantified by the Reynolds number $\textrm{Re}$ introduced in Section V.

\section{Resistance measurements and the leaky diode}

In Fig. \ref{fig:teslaplot}(a), we present as the green markers and curves measurements of flow rate $Q$ for varying height differential $\Delta h$. Here $\Delta h > 0$ corresponds to $Q>0$ or flow in the forward or `easy' direction (filled markers), while the reverse or `hard' direction corresponds to $\Delta h < 0$ and $Q<0$ (open markers). As might be expected, increasing the height differential yields higher magnitudes of flow rate in both cases. The absolute flow rate $|Q|$ increases monotonically but nonlinearly with $|\Delta h|$ for flow in both directions. More important but more subtle is that, for the same $|\Delta h|$, the values of $|Q|$ differ for forward versus reverse, the former being greater than the latter across all values $|\Delta h|$. This anisotropy is more clearly seen in Fig. \ref{fig:teslaplot}(b), where the resistance $R=\Delta p / Q$ is plotted versus $|Q|$ for the forward and reverse cases. Across all values of $|Q|$ explored here, the resistance in the reverse direction is higher than that of the forward direction.

\begin{figure}
\centering
\includegraphics[width=16.5cm]{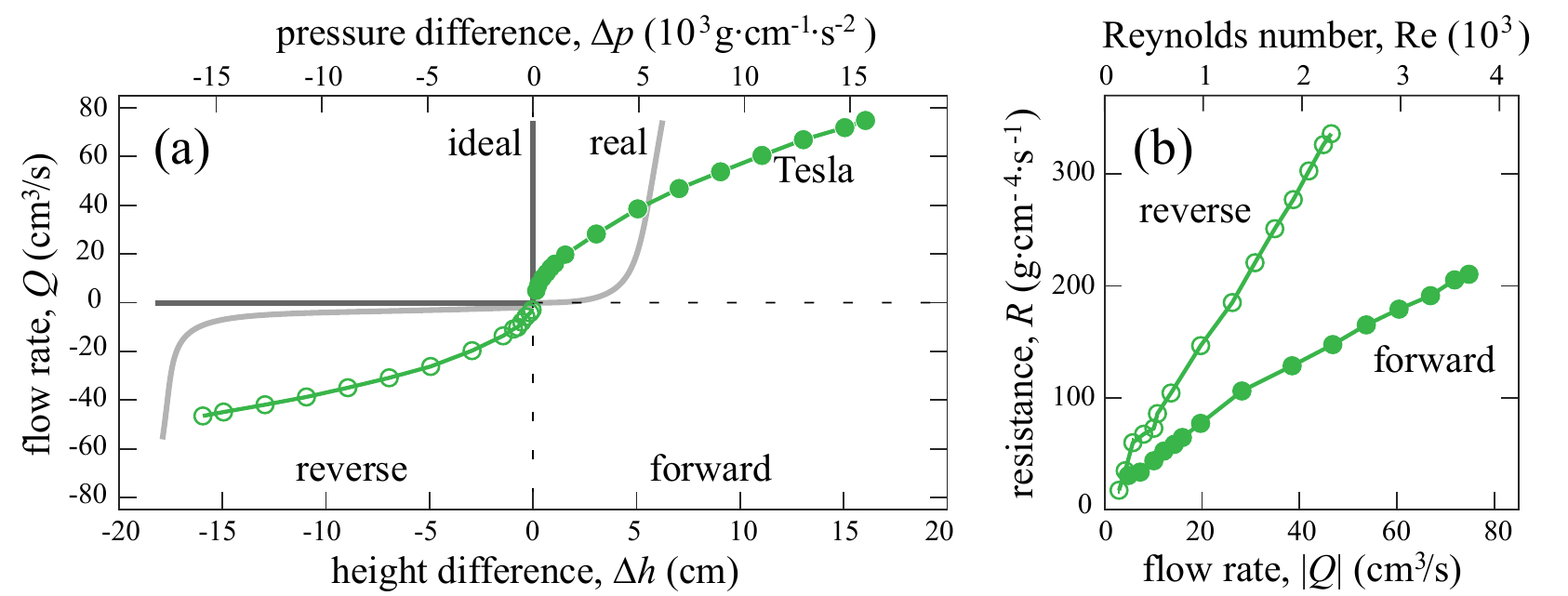}\vspace{-0.6cm}
\caption{Experimental tests of Tesla's conduit. (a) Green markers and curves indicate measured flow rate $Q$ across varying differences in water column heights $\Delta h$ and thus differential pressure $\Delta p$. The forward direction (filled symbols) permits greater flow rates for the same $\Delta h \propto \Delta p$. Curves are also shown for the behavior of an ideal diode (dark gray) and a typical real electronic diode (light gray). (b) Inferred resistance $R=\Delta p / Q$ versus $|Q|$ for flow in both directions. The conduit acts as a leaky diode with higher but finite resistance in the reverse direction. Error bars are smaller than the symbol size.}
\label{fig:teslaplot}
\end{figure}

A note about errors: The flow rate $Q$ is determined by triggering a stopwatch when a given volume of liquid is collected, and so errors are set by the human visual reaction time ($\sim 0.3$ s). Relative errors under 1\% are achieved simply with long collection times ($>30$ s). The differential height $\Delta h$ is determined by visually reading the water column heights on vertical rulers in each chamber, and so errors are set by the height of the meniscus ($\sim 1$ mm). For a typical $|\Delta h|$ of 10 cm, the errors are about 1\%. We suppress error bars in Fig. \ref{fig:teslaplot} and elsewhere when they are smaller than the symbol size.

The data of Fig. \ref{fig:teslaplot} provide direct experimental validation for Tesla's main claim of anisotropic resistance. But, at least for the conditions studied here, the ratio of hard to easy resistances is far less than that reported in Tesla's patent, being closer to 2 times rather than 200 times. This factor is quantified and compared against other channel designs in Section VIII, and we discuss possible reasons for this discrepancy in Section IX.

Returning to the electronic-hydraulic analogy, our results suggest that the conduit acts as a \textit{leaky diode}. To appreciate this term, it is useful to compare our results against the performance of ideal and real electronic diodes, as shown in Fig. \ref{fig:teslaplot}(a). An ideal diode offers no resistance in the forward direction and thus $Q$ is infinite for all $\Delta h > 0$. It has infinite resistance in reverse and thus $Q=0$ for all $\Delta h < 0$. A real electronic diode typically requires a finite voltage to `turn on' in the forward direction, has some small leakage current in reverse, and breaks down for very large reverse voltages. Our measurements indicate that Tesla's conduit deviates in all such features, but a common trait is the leakage in reverse, which is quite substantial for the conditions studied here.

These results can be summarized by the representation of the leaky diode as shown in Fig. \ref{fig:setup}(a). Symbolized as four arrowheads, it is equivalent to a parallel pair of resistors and ideal diodes, themselves arranged in series within each pair. Positive voltages drive current through a forward resistance $R_1$, and negative voltages drive lower current through a higher reverse resistance $R_2 > R_1$. The analogy is made more exact if the resistance values are functions of current.

\section{Irreversibility of high Reynolds number flows}

The fundamental characteristic of Tesla's device borne out by the above measurements is that, when the applied pressures are reversed, the flows do not simply follow suit by reversing as well. Rather, substantially different flow rates result, and presumably all details of the forward versus reverse flows through the conduit differ as well. This is a manifestation of \textit{irreversibility}, a property that arises more generally in many physical contexts.\cite{gollub2006,hollinger2012} Flow irreversibility was anticipated by Tesla, whose drawing reproduced in Fig. \ref{fig:patent}(c) shows a rather straight trajectory (dashed line and arrow) down the central corridor for the forward direction and a more circuitous route around the islands for the reverse direction. This outcome could be viewed as unsurprising; after all, the channel is clearly asymmetric or directional. For those new to fluid mechanics, it may be counter-intuitive that there exist conditions for which flows are exactly reversible even for asymmetric geometries, implying equality of the forward and reverse resistance values. (Tesla may not have been aware of this.) Fluid dynamical or kinematic (ir)reversibility -- which is distinct from the thermodynamic (ir)reversibility of a process -- can be derived from the governing Navier-Stokes equation of fluid dynamics, an analysis taken up elsewhere.\cite{stone2004,stone2017} Its central importance to the function of Tesla's device warrants a brief overview of known results.

While other forces may participate in various situations, three effects are intrinsic to fluid motion: pressure, inertia and viscosity. It is useful to think of flows as being generated by pressure differences overcoming the inertia of the dense medium and its viscous resistance. In well known results for laminar flows that can be found in fluid mechanics textbooks,\cite{tritton} the inertial pressure scales as $\rho U^2$ and the viscous stress as $\mu U/ \ell $, where $\rho$ is the fluid density, $\mu$ its viscosity, $U$ a typical velocity, and $ \ell $ a relevant length scale. The relative importance of inertia to viscosity can be assessed by their ratio, which is the dimensionless Reynolds number $\textrm{Re} = \rho U^2 / (\mu U / \ell ) = \rho U \ell  /\mu$. In the low Reynolds number regime of $\textrm{Re} \ll 1$, inertial effects can be ignored, and the resulting linear Stokes equation is reversible. \cite{tritton,stone2017} Qualitatively, viscosity causes flows to stick to solid boundaries and conform to identical paths for forward and reverse directions. This general property of viscous flows has many important consequences and is beautifully demonstrated by stirring a viscous fluid and then `unstirring' with precisely reversed motions, causing a dispersed dye to recollect into its original form. \cite{taylor1967,fonda2017} When Re is not small, inertial effects participate and flows are governed by the full Navier-Stokes equation, whose nonlinearity leads to irreversibility.\cite{tritton} Qualitatively, inertia allows flows to depart or separate from solid surfaces, this tendency being sensitive to geometry and thus directionality. Among many other phenomena, this relates to the observation that one can blow out but not suck out a candle, the flows being markedly different under the reversal of pressures.  

For computing Reynolds numbers for pipe flows, it is customary to set $U$ as the average flow speed and to use the diameter $D$ (or a corresponding dimension for non-circular conduits) as the length scale. Saving a deeper discussion of these quantities for Section VII, the parameters in our experiments yield $\textrm{Re} = \rho U D /\mu \sim 10^2-10^4$, as reported on the upper axis of Fig. \ref{fig:teslaplot}(b). Such high values of Re indicate that the flows are strongly inertial and thus irreversible, which is consistent with the different forward versus reverse resistance values reported here. The directional dependence of high-Re flows has been observed previously in computational fluid dynamics simulations for modified forms of Tesla's channel.\cite{bardell2000,nobakht,anag,zhang,thompson} For low Re, flow reversibility has been confirmed by experimental visualization,\cite{stone2017} and future work should verify the symmetric resistance expected in this regime.

\section{Analysis of pressures in two-chamber system}

In the above analysis of the experimental data, we have assumed that our two-chamber apparatus imposes $\Delta p = \rho g \Delta h$ across the channel. This formula strictly represents the gravitational \textit{hydrostatic} pressure difference between liquid columns, whereas the fluid is in motion throughout our system, so what justifies its application here? The short answer is that the flows in the chambers outside of the channel are ``slow enough'' to safely ignore velocity-dependent pressures. The long answer is provided in this section, where we analyze the contributing pressures in the system via Bernoulli's law. This analysis also sets up the following section by showing that Bernoulli's law is violated in the channel itself, requiring a characterization of friction or dissipation. 



Bernoulli's law is a statement of conservation of energy for steady flows of an inviscid (zero-viscosity) fluid. \cite{tritton} Of course, real fluids like water have finite viscosity, and later we discuss the validity of this approximation for the flows in the chambers. Following streamlines of the flow, the pressure, the gravitational potential energy density (\textit{i.e.} energy per unit volume), and the kinetic energy density must each change in a way such that their sum (total energy density) is unchanged: $H = p + \rho g z + \frac{1}{2} \rho U^2$ is constant. Here $p$ is the pressure, which can be thought of as a measure of internal energy density, $z$ is the vertical coordinate, $U$ is the speed at any location along a streamline, and $H$ is the total energy density. 

\begin{figure}
\centering
\includegraphics[width=8cm]{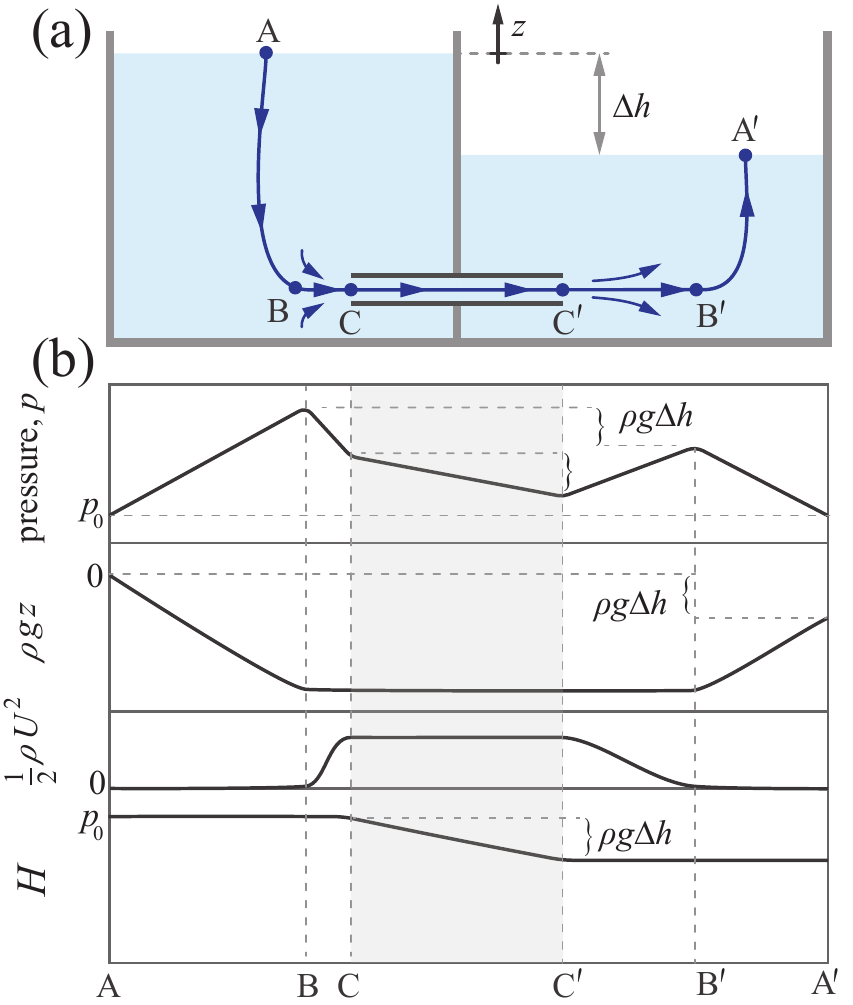}\vspace{-0.6cm}
\caption{An analysis of the two-chamber system guided by Bernoulli's law. (a) Chambers of high and low water levels are connected only by a channel at some depth. A flow streamline ABCC$'$B$'$A$'$ starts at the surface of the higher column, enters the channel, and emerges in the lower column where it meets the surface. (b) Hypothetical data showing how pressure $p$, gravitational potential energy density $\rho g z$, and flow kinetic energy density $\frac{1}{2} \rho U^2$ and total energy density $H= p + \rho g z + \frac{1}{2} \rho U^2$ vary along the streamline.}
\label{fig:bernoulli}
\end{figure}

In Fig. \ref{fig:bernoulli}(a) we examine a hypothetical streamline assumed to originate at the surface of the high-level chamber, descend to the channel opening, transit through the channel and out into the low-level chamber, and finally up to the surface. In its descent from the point marked A at the free surface to the point B somewhat upstream of the channel inlet, it is reasonable to assume that flow speed remains always small $U\approx 0$ and so too does the kinetic energy (more on this approximation later). In this case, the primary energy exchange involves a drop in the gravitational term $\rho g z$ and a consequent rise in $p$, as shown in the segment A-B in Fig. \ref{fig:bernoulli}(b), which tracks the terms of Bernoulli's law with hypothetical data.

A similar exchange occurs for the end segment B$'$-A$'$. Because the free surface points A and A$'$ are both at atmospheric pressure $p_0$, we conclude that pressure difference between points B and B$'$ is $\rho g \Delta h$. The regions B-to-C just before the inlet and C$'$-to-B$'$ just after the outlet are approximated as horizontal and so involve exchanges of pressure with kinetic energy only. The flow becomes faster from B-to-C as it is constricted and becomes slower C$'$-to-B$'$ as it spreads out, and these changes in speed $U$ must come with changes in pressure. However, the increase in speed from B-to-C would seem to be matched by the decrease from C$'$-to-B$'$, and so the pressure drop B-to-C is matched by the rise C$'$-to-B$'$. If true, then indeed the pressure drop across the channel C-to-C$'$ is $\Delta p = \rho g \Delta h$.

This conclusion rests on the assumptions that the fluid has negligibly small viscosity (so that Bernoulli's law may be used) and that the flow speeds outside of the channel are negligibly slow (so motion-dependent pressures may be ignored). Such approximations are statements about the \textit{relative} strengths of effects, which can be quantified by dimensionless numbers representing ratios of participating forces. \cite{tritton} For the flows in the chambers of our device, we have not only the intrinsic effects of fluid inertia or kinetic energy and viscous stresses or dissipation but also gravitational pressure or potential energy. Following up on the force scales introduced in the preceding section, the associated stresses or energy densities are $\rho U^2$, $\mu U/ \ell $ and $\rho g \ell $, respectively, where $\rho$ is the fluid density, $\mu$ its viscosity, $U$ a typical velocity, and $\ell $ a relevant length scale.\cite{tritton} Concerned only with orders of magnitude, $g \sim 10^3~ \textrm{cm}/\textrm{s}^2$, $\rho \sim 1~\textrm{g}/\textrm{cm}^3$ and $\mu = 10^{-2}~\textrm{dyn} \cdot \textrm{s} / \textrm{cm}^{2}$ for water, $\ell  \sim 10$ cm is the chamber size, and $U = Q/\ell ^2 \sim 0.1$ cm/s for the typical flow rates explored here. The gravitational-to-viscous ratio is then $\rho g \ell  / (\mu U / \ell ) = \rho g \ell ^2 / \mu U \sim 10^8 \gg 1$, which certainly justifies the neglect of viscosity. (For those familiar with dimensionless numbers in fluid mechanics, this ratio can be expressed in terms of the Galileo and Reynolds numbers.) The gravitational-to-kinetic ratio is $\rho g \ell  / \rho U^2 = g \ell  / U^2 \sim 10^6 \gg 1$, which justifies the neglect of kinetic effects. (This ratio is related to the Froude number.) In essence, the pressure is very nearly balanced by the gravitational hydrostatic pressure throughout the chambers, with all other effects being many orders of magnitude smaller.

\section{Characterizing fluid friction in Tesla's conduit}

The above reasoning about pressures in the system indicates that Bernoulli's law is violated in the channel itself: The total energy density $H = p + \rho g z + \frac{1}{2} \rho U^2$ shown in the lowest panel of Fig. \ref{fig:bernoulli}(b) is not constant over the gray region C to C$'$. The gravitational energy density $\rho g z$ is unchanged over the horizontal length, and the kinetic energy density $\frac{1}{2} \rho U^2$ is unchanged due to mass conversation and thus uniformity of flow speed, so pressure varies without any variation in potential or kinetic energy. This is expected since the flow inside the channel is resisted due to viscosity, which dissipates energy and may trigger turbulence or unsteady flows, effects which are not accounted for in Bernoulli's law. The associated hydraulic resistance or friction is, of course, well studied in the engineering literature due to its practical importance, and in this section we apply established characterizations to our measurements on Tesla's channel. We also compare our findings to previous results on rough-walled pipes, which may serve as a crude (rough?) way to view Tesla's channel.

\begin{figure}
\centering
\includegraphics[width=7cm]{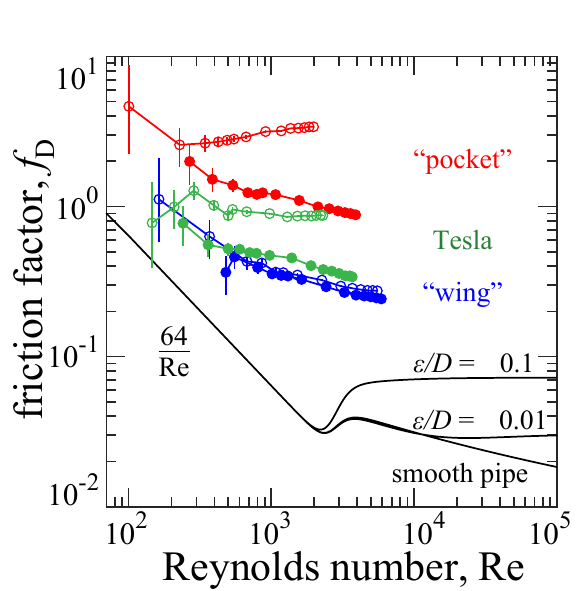}\vspace{-0.4cm}
\caption{Fluid friction factor versus Reynolds number (Moody diagram) for Tesla's channel and compared to smooth and rough pipes. Friction factors are plotted for forward and reverse flow through Tesla's channel (green) as well as two additional diode designs (red and blue) and circular pipes of different relative roughness $\varepsilon/D$ (black).}
\label{fig:moody}
\end{figure}


Hydraulic resistance depends on the conduit geometry as well as the Reynolds number, which assesses the relative importance of fluid inertia to viscosity. Following the discussions in previous sections, internal flows are characterized by $\textrm{Re} = \rho U D /\mu$, where $\rho$ and $\mu$ are the fluid density and viscosity, and $U$ is the average flow speed, and $D$ is the pipe diameter or a corresponding dimension in the case of non-circular cross-sections. For our realization of Tesla's channel, we use the so-called hydraulic diameter $D = 4V/S = 0.8$ cm, where $V$ is the total wetted volume of the conduit and $S$ is its total wetted surface area. (This is a generalization of the conventional form $D = 4A/P$ for a conduit whose cross-section shape is uniform and of wetted area $A$ and perimeter length $P$.\cite{white1999}) The average flow speed is $U = Q/A = 1 - 100 $ cm/s, where $A=wd$ is the average wetted area of the cross-section. These parameters yield $\textrm{Re} \sim 10^2-10^4$, as mentioned in Section V and reported on the upper axis of Fig. \ref{fig:teslaplot}(b).

A dimensionless measure of hydraulic resistance used often in engineering is the Darcy friction factor \cite{weisbach1845} $f_{\mathrm{D}} = (\Delta p/L)/(\frac{1}{2} \rho U^2/D  )$, where $L$ is the conduit length and $\Delta p$ is the pressure loss along the conduit or, equivalently for our set-up, the applied pressure difference. This choice of nondimensionalization has been shown to yield values of $f_{\mathrm{D}}$ that vary only weakly with Re for turbulent flow through long pipes.\cite{moody1944, bello2018} The green markers and curves of Fig. \ref{fig:moody} represent the measured friction factors versus Reynolds number for forward and reverse flow through Tesla's channel, and data are included for two additional diode designs that will be introduced in Section VIII. For comparison, we include the so-called Moody diagram,\cite{moody1944,bello2018} which is a log-log plot summarizing measurements of $f_{\mathrm{D}}(\textrm{Re})$ for circular pipes of varying degrees of wall roughness. Here the relative roughness $\varepsilon/D$ represents the ratio of typical surface deviations $\varepsilon$ to the mean diameter $D$. Smooth and rough pipes alike follow a well-known form of $f_{\mathrm{D}} = 64/\textrm{Re}$ in the laminar flow regime of $\textrm{Re} \lesssim 2 \times 10^3$. (This form can be derived from the Hagen-Poiseuille law for developed, laminar flow in cylindrical pipes. \cite{tritton}) At higher $\textrm{Re} \gtrsim 4 \times 10^3$, the flow tends to be turbulent, and $f_{\mathrm{D}}$ varies weakly with Re but increases with wall roughness.

Interestingly, the friction factors for flow through Tesla's conduit are far higher than those reported for smooth and rough pipes at comparable Re. This likely reflects the extreme degree of roughness ($\varepsilon/D \sim 1$, if such a quantity is at all meaningful) of the channel and consequent disturbances to the flow presented by its baffles and islands. It is clear that our results do not follow the form $f_{\mathrm{D}} = 64/\textrm{Re}$ for the range $\textrm{Re} \approx 200-2000$ explored here, nor is there any clear feature in the curves that would indicate a laminar-to-turbulent transition. Making sense of these observations, and better understanding the hydraulics of very rough channels generally,\cite{gloss2010, liu2019} would benefit from further work that varies Re over a wider range and includes flow visualization.

\section{Alternative diode designs and comparison of diodicity}

Equipped with the experimental methods and with a grasp of the basic fluid dynamics involved, we next pose as a challenge to design a fluidic diode that outperforms Tesla's valvular conduit. In principle, any channel with asymmetric internal geometry may have asymmetric resistance at appreciably high Re. However, designing a channel with high resistance ratio is a challenging shape optimization problem, and the electronic-hydraulic analogy is not informative when it comes to issues of detailed fluid-structure interactions. In the absence of any such well-informed strategy for ``intelligent design'', and without the patience for evolutionary algorithms that may iteratively and systematically improve the shape, we borrow Tesla's use of intuition and inspiration to arrive at the two alternative diodes shown in Figs. \ref{fig:twodiodes}(a) and (b), which we then construct and test.

\begin{figure}
\centering
\includegraphics[width=15cm]{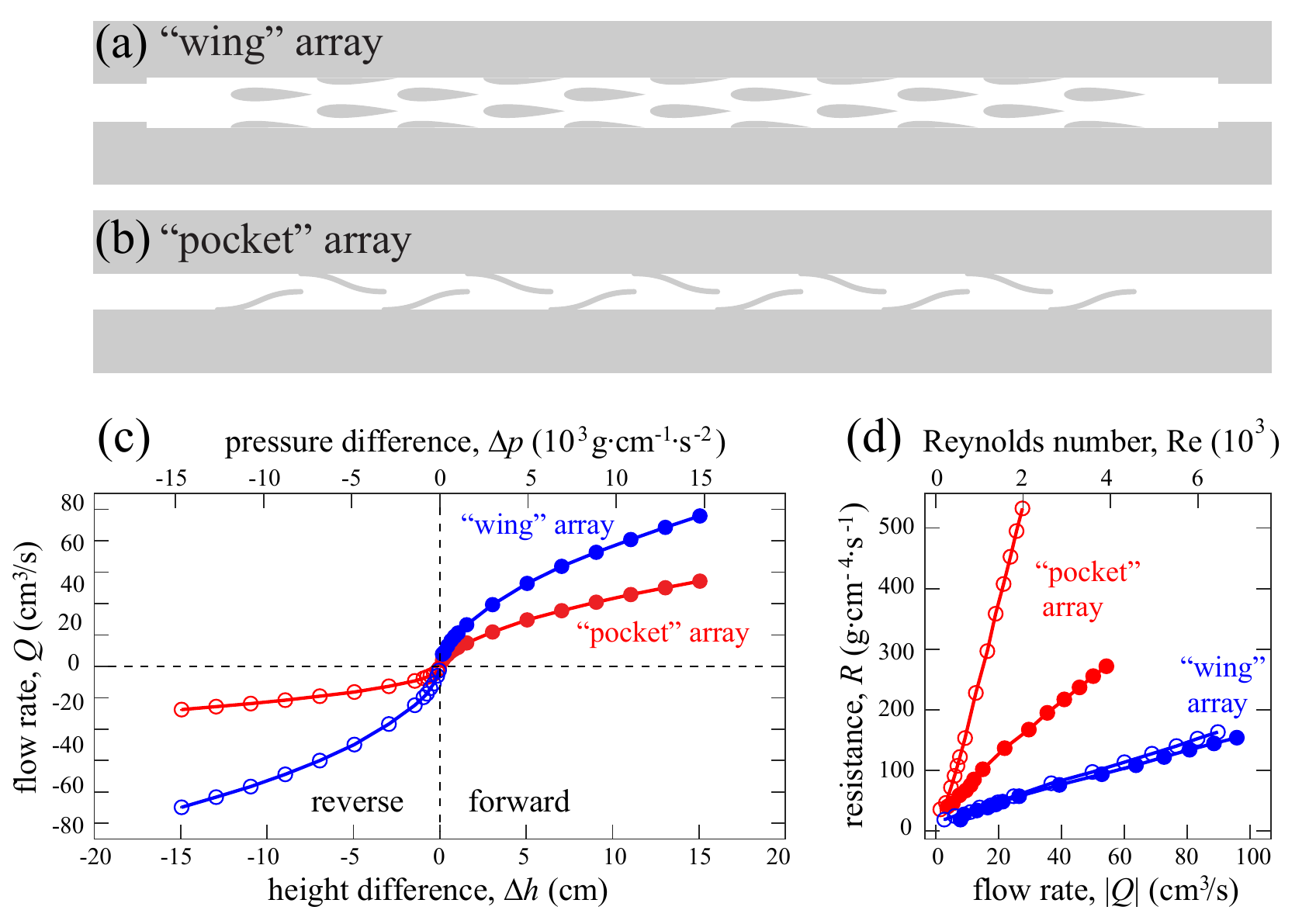}\vspace{-0.6cm}
\caption{Two alternative diode designs and measurements of their performance. (a) An array of wing or airfoil sections. The fluid is to occupy the white regions between the gray regions representing solid material. (b) An array of sigmoidal baffles forms dead-end `pockets' when viewed from the reverse direction. (c) Characterization of the designs through measurements of volume flow rate $Q$ versus water level height differential $\Delta h$ or differential pressure $\Delta p$. (d) Resistance $R = \Delta p / Q$ versus absolute flow rate $|Q|$ for forward and reverse operation of both designs.}
\label{fig:twodiodes}
\end{figure}

To facilitate direct comparison against Tesla's design, we imposed the following criteria on our channels: They must be periodic with the same number (11) of repeating units and the same length (30 cm), depth (1.9 cm), and average width (0.9 cm). The first design of Fig. \ref{fig:twodiodes}(a) employs a staggered array of wings or airfoil shapes whose rounded leading edges face into the flow in the forward operation of the diode. The reasoning is that wings, at least when used individually in their typical application of forward flight, are intentionally streamlined for low resistance. Flow in reverse, however, is can trigger flow separation near the thickest portion of the airfoil section and thus a wide wake. Our second design shown in Fig. \ref{fig:twodiodes}(b) replaces Tesla's `buckets', which reroute flows in the reverse mode, with dead-end `pockets' formed by sigmoid-shaped baffles. 

Repeating the experimental procedures outlined above, we arrive at curves for $Q(\Delta h)$ for both designs operating in both forward and reverse, as shown by the plots of Fig. \ref{fig:twodiodes}(c). The corresponding resistance curves $R(Q)$ are shown in Fig. \ref{fig:twodiodes}(d). Surprisingly, the wing design is nearly isotropic with the forward and reverse resistances almost equal across all flow rates tested. Perhaps the similarity in resistance values could be explained by the suppression of flow separation in both directions due to the confined geometry of the channel. In any case, it is clear that not all asymmetric geometries lead to strongly asymmetric resistances even for high Re flows. The pocket design fares better, with a resistance in reverse that is substantially higher than the forward resistance.

\begin{figure}
\centering
\includegraphics[width=10cm]{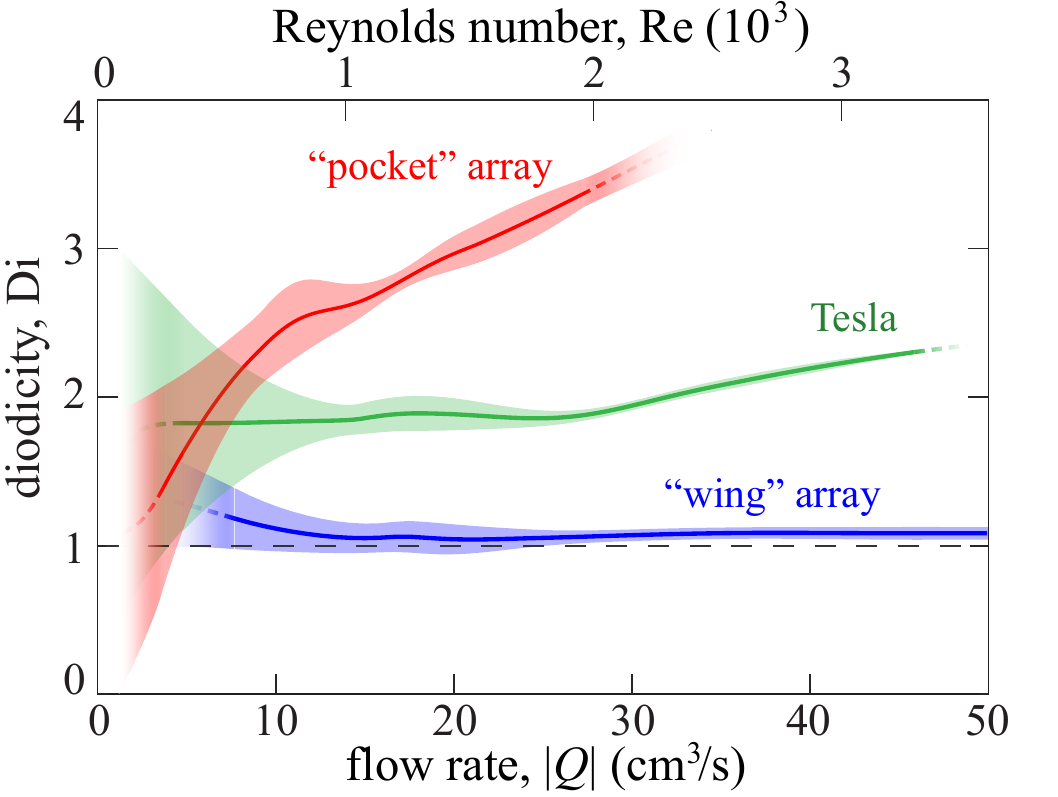}\vspace{-0.4cm}
\caption{Comparing the performance of three channel designs. The diodicity $\mathrm{Di}(|Q|) = R_2(|Q|)/R_1(|Q|)$ of a given channel is the ratio of reverse to forward resistances at the same magnitude of flow rate. The pocket-array design outperforms Tesla's conduit, while the wing-array is nearly symmetric in resistance.  Each $R(|Q|)$ is interpolated from $R$ versus $|Q| $ data. Shaded regions reflect standard error of the mean propagated from measurement errors of $Q$ and $\Delta h$.}
\label{fig:diodicity}
\end{figure}

To quantitatively compare the performance of all three diode designs, we define the ratio of reverse resistance $R_2$ to forward resistance $R_1$ as the \textit{diodicity},\cite{forster1995} $\mathrm{Di} = R_2/R_1$. Specifically, we may evaluate $\mathrm{Di}(|Q|) = R_2(|Q|)/R_1(|Q|) = \Delta p_2(|Q|) / \Delta p_1(|Q|) $, which is also equivalent to $\mathrm{Di}(\mathrm{Re}) = f_2(\textrm{Re})/f_1(\textrm{Re})$. Because $R_1$ and $R_2$ are not in general measured at the same $|Q|$, we fit curves to these data and compute their ratio, resulting in the plots shown for Tesla's design and our two diodes in Fig. \ref{fig:diodicity}. The shaded regions indicated errors propagated from the raw measurements. The diodicity of Tesla's conduit is a weakly increasing function of flow rate with a typical value of $\mathrm{Di} \approx 2$ for the conditions studied here. The wing-array design has weak diodicity near unity. Interestingly, the pocket-array design has a more strongly increasing $\mathrm{Di}(|Q|)$ curve than does Tesla's conduit, and it significantly outperforms Tesla's design over most of the range of $Q$ tested. If the trend continues to higher flow rates (higher Re), then even greater diodicity values can be expected.

\section{Discussion and conclusions}

Nikola Tesla's valvular conduit is an engaging context to introduce students of all levels to the role played by creativity in scientific research and specifically the power and limitations of reasoning by analogy. The information presented here can be used as a guide for lectures in an introductory fluid mechanics course, a module in a laboratory course, or as a springboard for a further research projects into fluidics and fluidic devices. As such, we have emphasized the practicalities of the experiments and their pedagogical value. The experimental apparatus can readily be made from household and standard laboratory items such as tanks, tubing and pumps, and the measurements may be accurately and efficiently carried out by students using rulers, beakers and stopwatches. The basic data analysis and plotting is rather straightforward, but at the same time there are ample opportunities for further analysis into errors and their propagation, curve fitting, dynamic resistance and differentiation of data, and so on. The fluid mechanics concepts of channel/pipe flow, hydraulic resistance/friction, Reynolds number, Bernoulli's law, low (ir)reversibility, etc. naturally arise from these investigations and our discussions are but brief introductions that may be followed up in depth. An additional project stems naturally from the challenge to design and test yet better diodes, in which case our pocket-array design might serve as the new standard to beat.

\begin{figure}
\centering
\includegraphics[width=13.5cm]{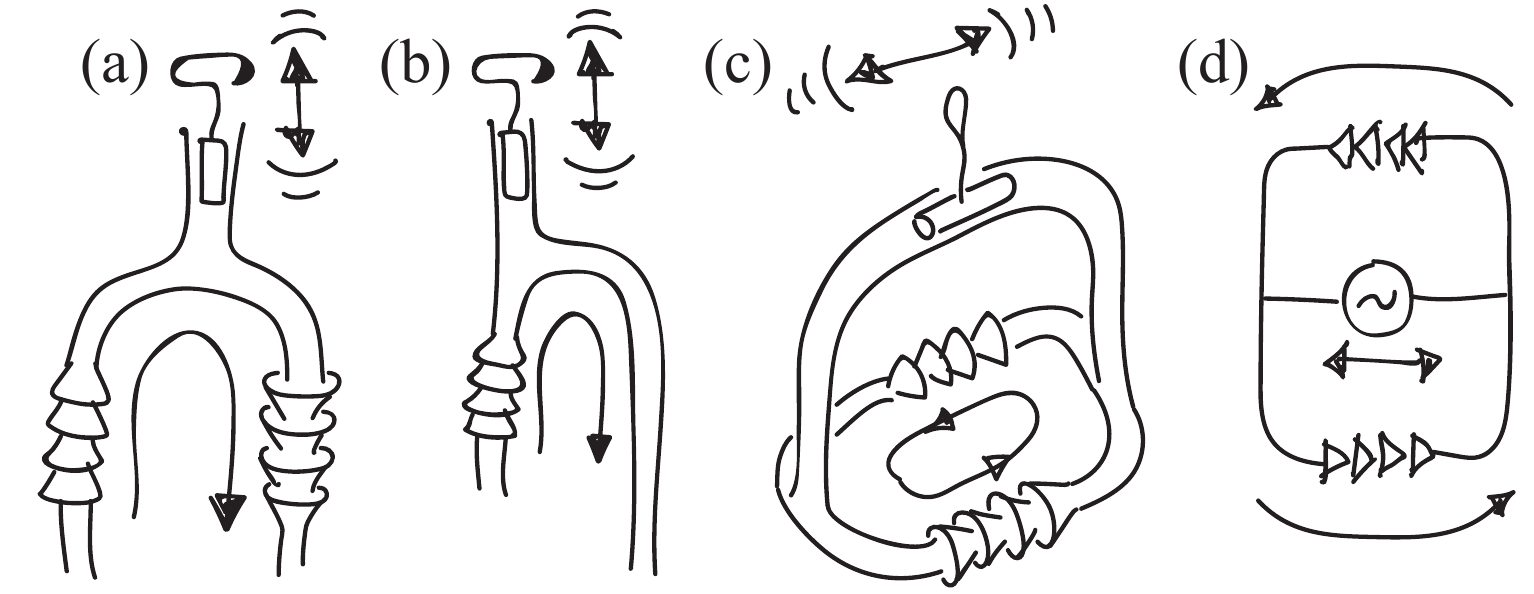}\vspace{-0.6cm}
\caption{Potential applications of fluidic diodes. (a) A hand pump in which oscillations of a piston are rectified by diodes to move fluid from one container to another. (b) A pump that uses a single diode to `prime' or fill a tube, which then drains a container by siphon action. (c) A closed system whereby oscillations drive one-way circulation. (d) An equivalent electronic circuit with AC source for the application shown in (c).}
\label{fig:applications}
\end{figure}

Another idea for an additional lesson plan or further direction for research involves the use of fluidic diodes in practical applications. In Fig. \ref{fig:applications}(a) we sketch a type of hand pump that might be used to transfer liquid from one container to another. The device is a bifurcating tube with a well-sealed piston on the upper branch and two diodes oppositely directed within the lower branches. Reciprocating motions of the piston may drive liquid up one branch and down the other. A related version with a single diode might exploit the rectification effect to `prime' or fill the tube, which then drains as a siphon if the outlet is held below a tank to be emptied [Fig. \ref{fig:applications}(b)]. An analogous but closed system is shown in Fig. \ref{fig:applications}(c). This represents a kind of circulatory system in which oscillations are rectified to drive flow around a loop. This could be used to pump coolant, fuel, lubricant or any of the other fluids that must be moved within machinery. The equivalent electronic circuit of Fig. \ref{fig:applications}(d) involves an alternating current (AC) source transformed into direct current (DC) by the diodes, the whole circuit acting as an AC-to-DC converter or rectifier.

It is also left for future work to explain the source of the large discrepancy between our measured diodicity of about 2 for Tesla's conduit and the `theoretical' and `approximate' value of 200 stated in his patent.\cite{patent} In any case, it is a worthwhile goal to enhance diodicity, which could involve modifying not only the conduit geometry but also the form of the imposed pressures and fluid properties beyond density and viscosity. Describing the reverse mode, Tesla conjectured that unsteady motions could be advantageous:\cite{patent} ``[T]he resistance offered to the passage of the medium will be considerable even if it be under constant pressure, but the impediments will be of full effect only when it is supplied in pulses and, more especially, when the same are extremely sudden and of high frequency.'' He may also have conceived of using air as the working fluid, for which compressibility effects would become important at extremely high speeds and pressures.

\begin{acknowledgments}
We thank S. Childress, M. Shelley and J. Zhang for useful discussions. We acknowledge support from the National Science Foundation (DMS-1646339 and DMS-1847955 to L.R.) and from the New York University Girls in Science, Technology, Engineering and Math (NYU GSTEM) program. 
\end{acknowledgments}

\newpage   
\setcounter{figure}{0}  
\begin{figure}
\centering
\caption{Nikola Tesla's valvular conduit. (a) The inventor Nikola Tesla (1856-1943). (b) Title to Tesla's 1920 patent for the valvular conduit \cite{patent}. (c) Schematic showing the conduit's internal geometry. The channel is intended to allow fluid to flow from left to right (forward direction) with minimal resistance while providing large resistance to reverse flow. }
\label{fig:patent}
\end{figure}

\begin{figure}
\centering
\caption{Experimental apparatus for testing Tesla's diode. (a) Circuit analogs. The resistance of an unknown component is characterized by imposing a voltage difference and measuring current. A `leaky' diode may be represented in terms of resistors and ideal diodes. (b) Rendering of Tesla's conduit that can be 3D printed or laser cut. Relevant dimensions include total length $L$, depth or height $d$, and average wetted width $w$ (c) Perspective view of experimental apparatus. Two upper chambers are connected only via the conduit to be tested, and both overflow through internal drains to a lower reservoir. The water level of the higher is maintained by an overspill mechanism with a pump. Flow through the channel into the lower chamber induces overflow into the reservoir. (d) Schematic of pressure source chamber. The height difference $\Delta h$ is set by variable-height drains, and the resulting volume flow rate $Q$ is measured. }
\label{fig:setup}
\end{figure}

\begin{figure}
\centering
\caption{Experimental tests of Tesla's conduit. (a) Green markers and curves indicate measured flow rate $Q$ across varying differences in water column heights $\Delta h$ and thus differential pressure $\Delta p$. The forward direction (filled symbols) permits greater flow rates for the same $\Delta h \propto \Delta p$. Curves are also shown for the behavior of an ideal diode (dark gray) and a typical real electronic diode (light gray). (b) Inferred resistance $R=\Delta p / Q$ versus $|Q|$ for flow in both directions. The conduit acts as a leaky diode with higher but finite resistance in the reverse direction. Error bars are smaller than the symbol size.}
\label{fig:teslaplot}
\end{figure}

\begin{figure}
\centering
\caption{An analysis of the two-chamber system guided by Bernoulli's law. (a) Chambers of high and low water levels are connected only by a channel at some depth. A flow streamline ABCC$'$B$'$A$'$ starts at the surface of the higher column, enters the channel, and emerges in the lower column where it meets the surface. (b) Hypothetical data showing how pressure $p$, gravitational potential energy density $\rho g z$, and flow kinetic energy density $\frac{1}{2} \rho U^2$ and total energy density $H= p + \rho g z + \frac{1}{2} \rho U^2$ vary along the streamline.}
\label{fig:bernoulli}
\end{figure}

\begin{figure}
\centering
\caption{Fluid friction factor versus Reynolds number (Moody diagram) for Tesla's channel and compared to smooth and rough pipes. Friction factors are plotted for forward and reverse flow through Tesla's channel (green) as well as two additional diode designs (red and blue) and circular pipes of different relative roughness $\varepsilon/D$ (black).}
\label{fig:moody}
\end{figure}

\begin{figure}
\centering
\caption{Two alternative diode designs and measurements of their performance. (a) An array of wing or airfoil sections. The fluid is to occupy the white regions between the gray regions representing solid material. (b) An array of sigmoidal baffles forms dead-end `pockets' when viewed from the reverse direction. (c) Characterization of the designs through measurements of volume flow rate $Q$ versus water level height differential $\Delta h$ or differential pressure $\Delta p$. (d) Resistance $R = \Delta p / Q$ versus absolute flow rate $|Q|$ for forward and reverse operation of both designs.}
\label{fig:twodiodes}
\end{figure}

\begin{figure}
\centering
\caption{Comparing the performance of three channel designs. The diodicity $\mathrm{Di}(|Q|) = R_2(|Q|)/R_1(|Q|)$ of a given channel is the ratio of reverse to forward resistances at the same magnitude of flow rate. The pocket-array design outperforms Tesla's conduit, while the wing-array is nearly symmetric in resistance.  Each $R(|Q|)$ is interpolated from $R$ versus $|Q| $ data. Shaded regions reflect standard error of the mean propagated from measurement errors of $Q$ and $\Delta h$.}
\label{fig:diodicity}
\end{figure}

\begin{figure}
\centering
\caption{Potential applications of fluidic diodes. (a) A hand pump in which oscillations of a piston are rectified by diodes to move fluid from one container to another. (b) A pump that uses a single diode to `prime' or fill a tube, which then drains a container by siphon action. (c) A closed system whereby oscillations drive one-way circulation. (d) An equivalent electronic circuit with AC source for the application shown in (c).}
\label{fig:applications}
\end{figure}

\end{document}